\documentclass[prb,twocolumn,showpacs,floatfix]{revtex4}
\usepackage{graphicx}% Include figure files
\usepackage{dcolumn}% Align table columns on decimal point
\usepackage{bm}% bold math
\begin{document}
\title{Direct observation of low energy nuclear spin excitations in HoCrO$_3$ by high resolution neutron spectroscopy}
\author{T. Chatterji$^1$, N. Jalarvo$^{2,3}$, C.M.N. Kumar$^{2,3}$, Y. Xiao$^3$ and Th. Br\"uckel$^3$ }
\address{$^1$Institut Laue-Langevin, 6 rue Joules Horowitz, BP 156, 38042 Grenoble Cedex 9, France\\
$^2$JCNS Outstation at the Spallation Neutron Source, Oak Ridge National Laboratory, Oak Ridge, Tennessee 37831-6475, USA\\
$^3$J\"ulich Centre for Neutron Science, Forschungszentrum J\"ulich D-52425 J\"ulich, Germany
}
\date{\today}

\begin{abstract}
We have investigated low energy nuclear spin excitations in strongly correlated electron compound HoCrO$_3$. We observe clear inelastic peaks at  $E = 22.18 \pm 0.04$ $\mu eV$ in both energy loss and gain sides. The energy of the inelastic peaks remains constant in the temperature range 1.5 - 40 K at which they are observed. The intensity of the inelastic peak increases at first with increasing temperature and then decreases at higher temperatures. The temperature dependence of the energy and intensity of the inelastic peaks is very unusual compared to that observed in other Nd, Co and V compounds.  Huge quasielastic scattering appears at higher temperatures presumably due to the fluctuating electronic moments of the Ho ions that get increasingly disordered at higher temperatures. 
\end{abstract}
\maketitle
The coupling of the nuclear spin system with the electronic spin system in condensed matter through hyperfine interaction  had been known for a rather long time but recently this phenomenon has attracted renewed interest due to its possible application in spintronics and quantum computation \cite{smet02,kane98}. 
 
    The principle of the method of studying the hyperfine interaction by the inelastic neutron scattering
is well known \cite{heidemann70,heidemann72} and will not be repeated. This method can be used to study the magnetic ordering by measuring the hyperfine splitting of the nuclear levels for magnetic materials that contain magnetic atoms with non-zero nuclear spins and also having large spin dependent neutron scattering cross sections. 
\begin{figure}
\resizebox{0.5\textwidth}{!}{\includegraphics{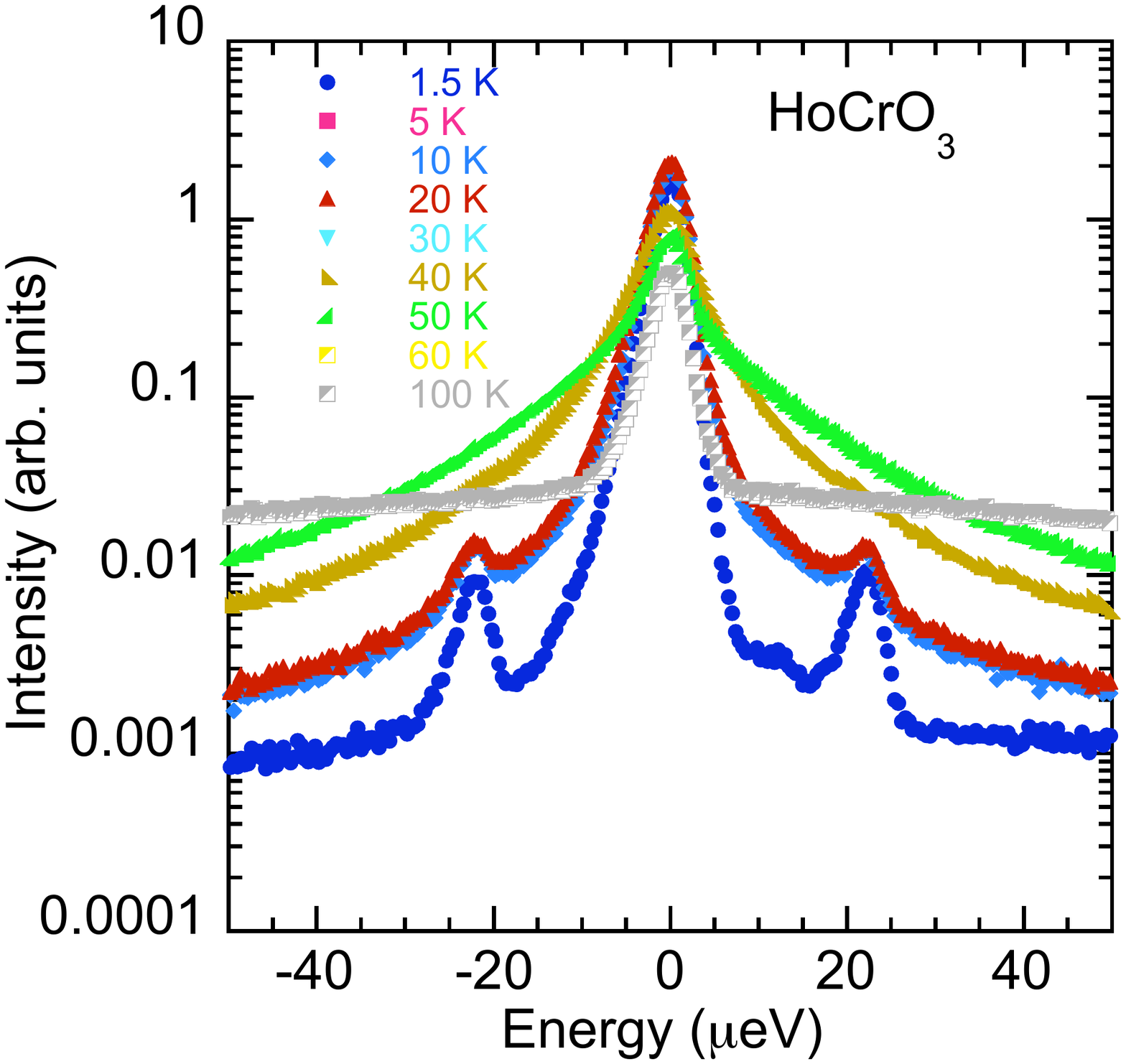}}

\caption { Q-integrated energy spectra of HoCrO$_3$ at several temperatures. 
           } 
 \label{hocrospectra}
\end{figure}

So far the high resolution inelastic neutron scattering technique has been used to study hyperfine interaction in Nd, Co, and V compounds \cite{heidemann72,chatterji00,chatterji02,chatterji04,chatterji04a,chatterji08,chatterji08a,chatterji09,chatterji11,chatterji09d,chatterji10} because their relatively large spin dependent scattering cross-sections. Ho compounds are also good candidates for such study since Ho has only one single stable isotope, $^{165}$Ho with a nuclear spin $I = 7/2$ and large magnetic dipole moment  \cite{dankwort74} of $4.17 \mu_N$. The neutron incoherent scattering cross section \cite{sears99} of  $^{165}$Ho is $0.36 \pm 0.03$ barn. Ehlers et al. \cite{ehlers09} observed recently nuclear spin excitations in Ho pyrochlore compound Ho$_2$Ti$_2$O$_7$. Here we report the results of our high-resolution inelastic neutron scattering investigation of the strongly correlated electron compound HoCrO$_3$. Recently rare earth cromites have drawn a considerable interest for their potential multiferroic behavior and application in spintronics. Our present investigation of the hyperfine interaction shades some light on a new control parameter, viz. the nuclear spin. The nuclear spin is coupled to the electric spin through hyperfine interaction. The electronic spin is coupled strongly to the lattice in multiferroic materials and gives rise to ferroelectric behavior. Thus nuclear spin can be available in this coupling process leading to the possibility of some novel application to spintronics and quantum computations. 

HoCrO$_3$ crystallizes with the orthorhombically distorted perovskite structure in the space group $D_{2h}^{16} - Pbnm$ and has four formula units per unit cell. The magnetic properties of HoCrO$_3$ have been studied by magnetization and neutron diffraction in the second half of the last century.  The exchange coupling between nearest-neighbor Cr$^{3+}$ ions is predominantly antiferromagnetic and these ions order magnetically \cite{bertaut66,pataud70,hornreich72} below $T_N \approx 140$ K. Below this temperature HoCrO$_3$ exhibits weak ferromagnetic moment \cite{hornreich72}. The magnetic structure of the Cr sublattice is $G_zF_x$ \cite{hornreich72} in the notation of Koehler et al. \cite{koehler60} and Bertaut \cite{bertaut63}. At lower temperature Ho ions order and due to strong interaction between Cr and Ho moments spin reorientation phenomena take place \cite{shamir77,kumar11}.The recent interest \cite{su11,lal96,park00,ramesha07,serrao05} in HoCrO$_3$ and other rare-earth cromites is due to the possibility of coupling between the magnetic and ferroelectric properties leading multiferroic behavior and their potential device applications.

Polycrystalline HoCrO$_3$ samples were synthesized by the solid state reaction of Ho$_2$O$_3$(3N) and Cr$_2$O$_3$(4N) in stoichiometric ratio \cite{kumar11}. The precursors were mixed intimately and subsquently heat treated at 1100 $^\circ$C for 48 h. The resulting material was then reground and annealed again at 1200 $^\circ$C for 24 h. The phase purity of the synthesized powder sample was then confirmed by powder X-ray diffraction with CuK$_\alpha$ ($\lambda = 1.54059$ {\AA}) radiation, using a Huber X-ray diffractometer (Huber G670) in transmission Guinier geometry. The powder is then pressed into pellets and sintered at 1000 $^\circ$C for 10 h for further magnetic and thermal characterization. Magnetic measurements were done on pressed-sintered pellets in a commercial (Quantum Design) superconducting quantum interference device magnetometer (SQUID) in the temperature range 2 - 300 K. Specific heat measurements were performed using a physical property measurement system (Quantum Design) in the temperature rage 0.1 - 290 K. The detailed results of the characterization of HoCrO$_3$ will be published elsewhere \cite{kumar12}.

Inelastic neutron scattering measurements were carried out on the BASIS back-scattering spectrometer \cite{mamontov11} of the Spallation Neutron Source (SNS) of the Oak Ridge National Laboratory in USA. 5.2 g of powder HoCrO$_3$ sample was first evenly distributed inside an Al foil and then placed inside the annular space of a cylindrical doubled wall Al sample holder. The sample holder filled with HoCrO$_3$ sample was placed inside the standard $^4$He cryostat.  

\begin{figure}
\resizebox{0.5\textwidth}{!}{\includegraphics{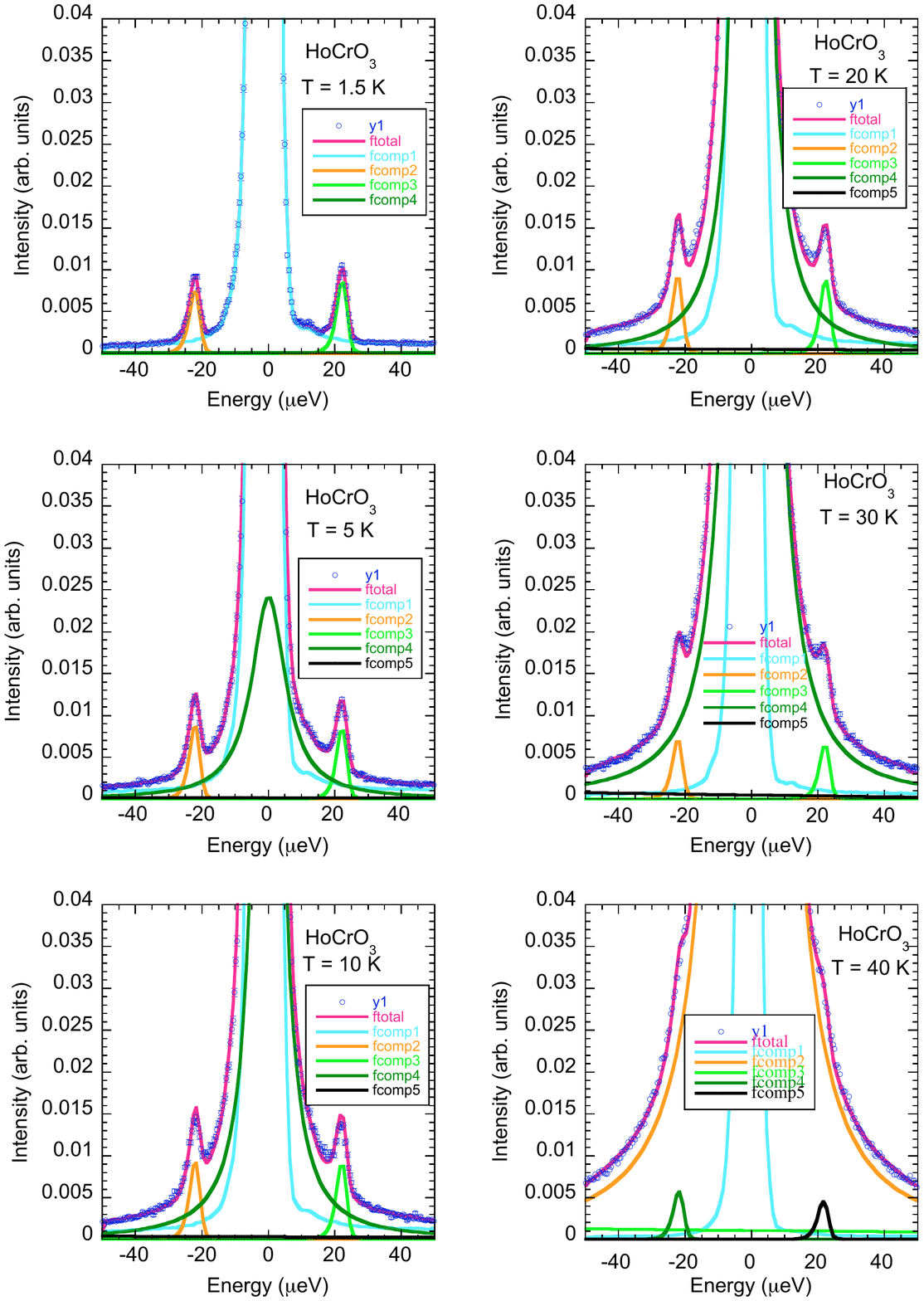}}

\caption { Inelastic peaks of HoCrO$_3$ at T = 1.5, 5, 10, 20, 30 and 40 K at both energy loss and gain sides fitted by convoluting instrumental resolution function determined from vanadium with two delta functions for the two inelastic peaks, one delta function for the elastic peak and a Lorentzian function for the quasielastic scattering. The different components are shown by continuous curves of different colors.
           } 
 \label{inelasticfit}
\end{figure}

\begin{figure}
\resizebox{0.5\textwidth}{!}{\includegraphics{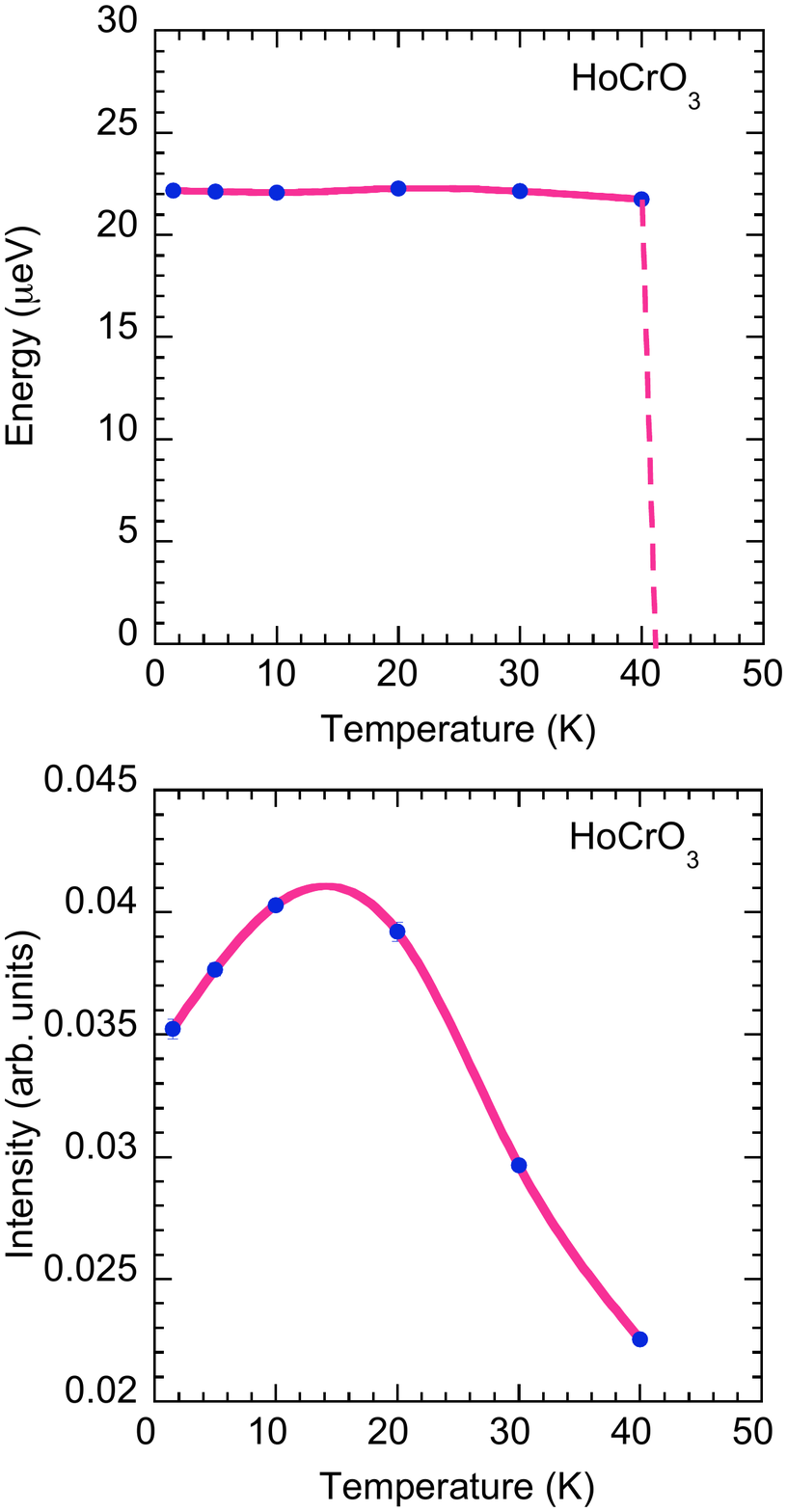}}
\caption {Temperature variation of the energy and intensity of the inelastic peak 
          of HoCrO$_3$. The continuous curves are just  guides to the eye. } 
\label{EI-Tdep}
\end{figure}

\begin{figure}
\resizebox{0.5\textwidth}{!}{\includegraphics{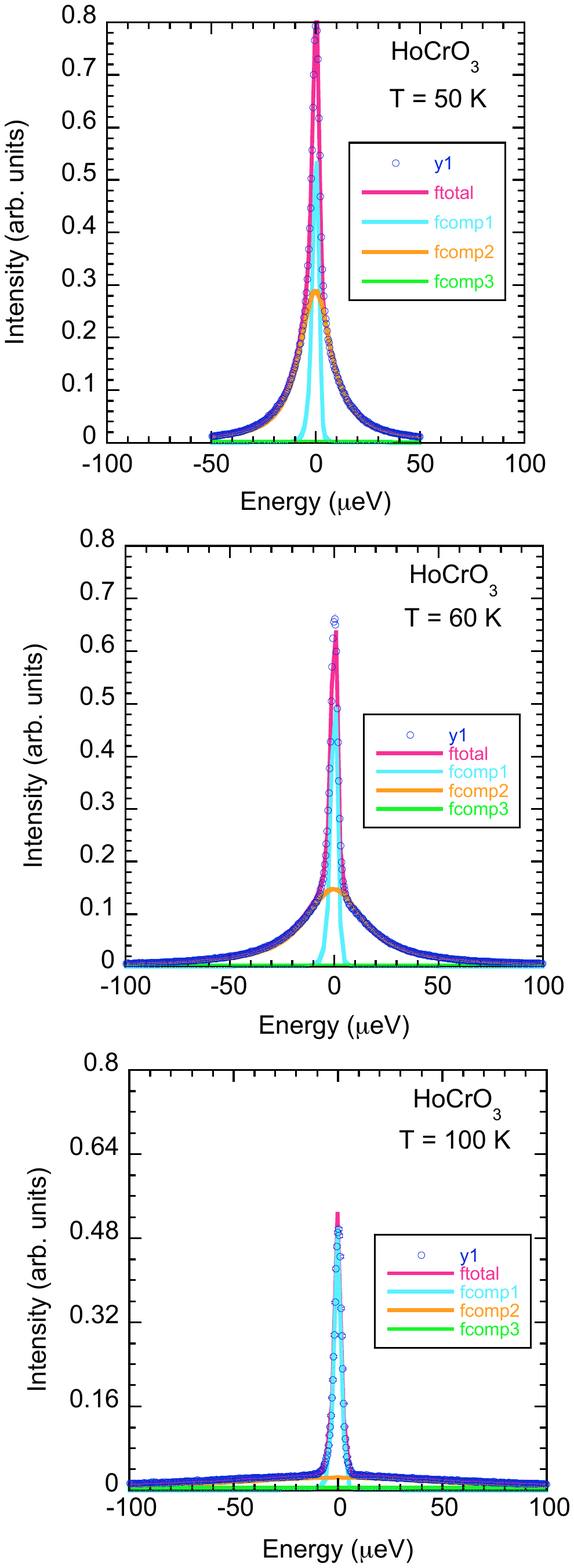}}

\caption { Quasielastic  scattering of HoCrO$_3$ at T =  50, 60 and 100 K fitted by convoluting the instrumental resolution function with a Lorenzian function.The different components are shown by continuous curves of different colors.
           } 
 \label{quasielasticfit}
\end{figure}

\begin{figure}
\resizebox{0.5\textwidth}{!}{\includegraphics{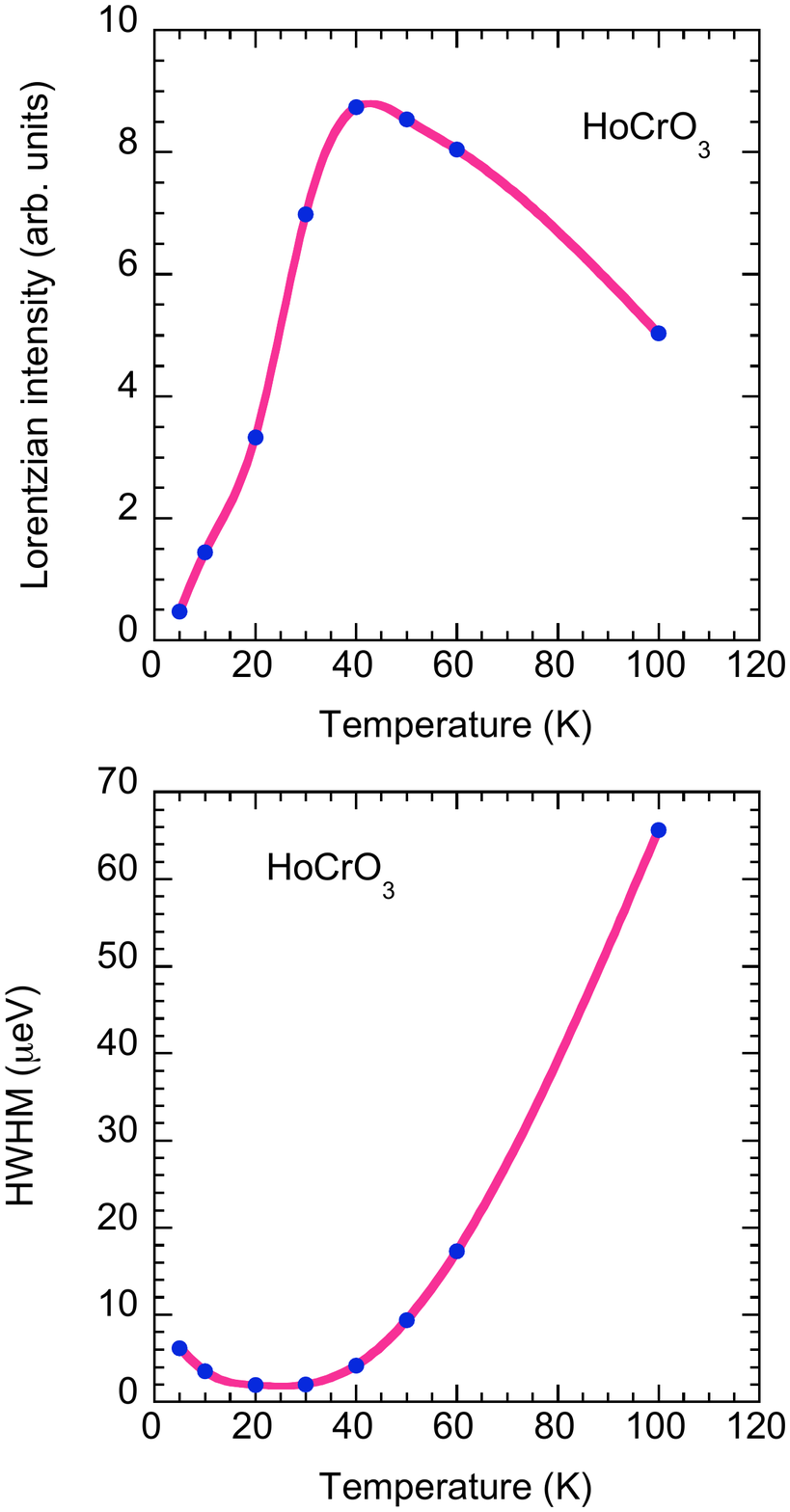}}
\caption {Temperature variation of the energy and intensity and half-width at half-maximum (HWHM) of the quasielastic scattering of HoCrO$_3$. The continuous curves are just  guides to the eye.} 
\label{IHWHM-Tdep}
\end{figure}

Fig. \ref{hocrospectra} shows the inelastic neutron scattering spectra of HoCrO$_3$ at several teperatures. At T = 1.5 K two clear inelastic signals are observed  on both energy gain and loss sides. The average energy of these inelastic signals is at $E = 22.18 \pm 0.04$ $\mu$eV which is very close to he hyperfine splitting $20.68 \mu$eV determined from the low temperature specific heat data \cite{kumar11}. We interpret these inelastic peaks to be due to the transition of the hyperfine-split nuclear levels. This energy is also very close to the energy (26.3 $\mu$eV) observed by Ehlers \cite{ehlers09} et al in the Ho pyroclore compound Ho$_2$Ti$_2$O$_7$.  We have fitted the spectra by the equation
\begin{eqnarray}
\label{fit}
S(\omega)=[x \delta_{el}(\omega)+p_1 \delta_{ins}(\omega)+p_2 \delta_{ins} (\omega) \\ \nonumber
+p_3L(\Gamma,\omega)] \otimes R(\omega)+B
\end{eqnarray}
where delta functions $\delta_{el}$ and $\delta_{ins}$ represent elastic and inelastic peaks respectively. The Lorentzian function $L$ represents the quasielastic component, $R(\omega)$ is the experimentally determined resolution function, B is the flat background term and $x, p_1, p_2, p_3$ are scaling factors. Fig. \ref{inelasticfit} shows the fit of  equation (\ref{fit}) to the measured data in several temperature in the range 1.5 to 40 K.
We note that the energy of inelastic peaks remains almost the same in the temperature range from 1.5 K to 40 K in which the inelastic peaks are visible. Fig. \ref{EI-Tdep} shows the temperature variation of the energy and intensity of the inelastic peak obtained from the least-squares fit. The energy of the inelastic peak remains almost constant in the temperature range 1.5 - 40 K and then becomes invisible.   This temperature variation of the energy of the nuclear spin excitations is quite different from that of Nd and Co compounds studied by us but is very similar to the Ho compound Ho$_2$Ti$_2$O$_7$ studied by Ehlers et al. \cite{ehlers09}. For Nd ad Co compounds studied by us the temperature dependence of the energy of nuclear spin excitations follows the order parameter of the magnetic phase transition, i.e., the energy decreases and becomes zero at the phase transition. The inelastic peaks move towards the central elastic peak and merge with it at the  magnetic phase transition. It is surprising that the energy of the inelastic peak does not follow the temperature dependence of the ordered magnetic moment of Ho ions deduced from neutron diffraction \cite{kumar11}. The intensity of the inelastic peak increases a bit with temperature and then decreases continuously and becomes very small at about 40 K. At higher temperatures the inelastic signal disappears or becomes invisible probably due to the appearance of huge quasielastic scattering. This behavior is also very different from the intensity variation observed in the Nd and Co compounds, but is again similar to that observed by Ehlers et al.  \cite{ehlers09} in Ho$_2$Ti$_2$O$_7$.

Another interesting outcome of the present study is the discovery of huge quasielastic scattering at higher temperatures. Fig. \ref{quasielasticfit} shows the spectra of HoCrO$_3$ measured at 50, 60 and 100 K.  At these temperatures the inelastic peaks due to hyperfine interaction are not observed. Strong quasielastic scattering dominates the spectra. We have fitted the spectra by the equation
\begin{equation}
\label{fit1}
S(\omega)=[x \delta_{el}(\omega)+
p_3L(\Gamma,\omega)] \otimes R(\omega)+B
\end{equation}
where the second and the third terms within square brackets of the previous equation (\ref{fit}) representing the two inelastic peaks have diappeared.
Fig. \ref{IHWHM-Tdep} shows the temperature variation of the intensity of the quasielastic scattering and its half-width at half-maximum (HWHM) of the least squares fit. The Lorentzian width gives the time scale of fluctuating magnetic moments. We interpret the quasielastic scattering to be due to the fluctuating disordered Ho electronic magnetic moments.  The large electronic magnetic moment of the Ho ions which of the order of about 10 $\mu_B$  can only give rise to such huge amount of quasielastic scattering. At low temperature the intensity of the quasielastic scattering is small because of the magnetic ordering of the Ho ions. At higher temperatures the magnetic moments become gradually disordered and therefore the fluctuations increase. The decreasing intensity at higher temperature may only indicate that the quasielastic scattering goes out of the energy window of the instrument and is therefore only partially integrated.  The half-width at half-maximum (HWHM) decreases at first with increasing temperature and then increases rapidly at higher temperatures. The temperature variation of HWHM is not yet quite clear. Ehlers et al.  \cite{ehlers09} have also observed huge quasielastic scattering in Ho$_2$Ti$_2$O$_7$ and have interpreted it to be due to the fluctuating electronic moments of the Ho ions but have not fitted their data. 

In conclusion we have observed low energy nuclear spin excitations in HoCrO$_3$ and have found quite different temperature dependence of the energy and intensity compared to those of Co and Nd compounds. We have also observed huge quasielastic scattering due to the fluctuating electronic moments of the Ho ions  that get increasingly disordered at higher temperatures.

\end{document}